\documentclass[draftcls,onecolumn,conference,12pt]{IEEEtran}

\usepackage{etoolbox}
\newtoggle{onecolumnformat}
\togglefalse{onecolumnformat}
\newcommand{\vect}[1]{{\mathbf{#1}}}
\newcommand{\norm}[1]{\| #1 \|}

\def\booknames#1/#2/#3/#4{#1:#2:#3:#4}
\usepackage{amsmath}
\allowdisplaybreaks
\usepackage{amsthm}
\usepackage[nothing]{algorithm}
\usepackage{algorithmic}
\usepackage{stfloats}
\usepackage{graphicx}
\usepackage{epsfig}
\graphicspath{ {figures/} }

\usepackage{dsfont}
\usepackage{graphicx}
\usepackage{epsfig}
\usepackage{bbm}
\usepackage{dsfont}
\usepackage{amsfonts}
\usepackage{array}

\usepackage{svn}
\SVN $Author: diego $ \SVN $Date: 2011-11-01 13:58:32 -0400 (mar.,
01 nov. 2011) $ \SVN $Rev: 1090 $

\usepackage[belowskip=-15pt,aboveskip=0pt]{caption}
\iftoggle{onecolumnformat} {\author{\IEEEauthorblockN{Diego
Perea-Vega, Jean-Fran\c cois Frigon, and Andr\'e Girard}
\IEEEauthorblockA{GERAD Research Center,
Electrical Engineering Department, \'Ecole Polytechnique de Montr\'eal\\
P.O. Box 6079, succ. Centre-Ville, Montr\'eal (Qu\'ebec), Canada, H3C 3A7.\\
Email: \{enrique.perea,j-f.frigon\}@polymtl.ca,
andre.girard@gerad.ca}}}
{\author{Diego Perea-Vega,~\IEEEmembership{Member,~IEEE}, Jean-Fran\c{c}ois Frigon,~\IEEEmembership{Member,~IEEE}, and Andr\'e Girard,~\IEEEmembership{Member,~IEEE}%
 \thanks{The authors are with the GERAD Research Center, Department of Electrical Engineering, \'{E}cole Polytechnique de Montr\'{e}al, Montr\'{e}al (Qc), Canada, H3T 1J4. e-mail: \{enrique.perea,j-f.frigon\}@polymtl.ca,
andre.girard@gerad.ca. This research project was supported by
NSERC under the Grant CRDPJ 335934-06.}}}

\DeclareMathOperator{\tr}{tr}

\DeclareMathOperator{\diag}{diag}

\begin{document}
%
\title{Efficient Heuristic for Resource Allocation in Zero-forcing MISO LTE-Advanced Systems with Minimum
Rate Constraints}
\maketitle
%
\begin{abstract}
4G wireless access systems require high spectral efficiency to
support the ever increasing number of users and data rates for
real time applications. Multi-antenna OFDM-SDMA systems can
provide the required high spectral efficiency and dynamic usage of
the channel, but the resource allocation process becomes extremely
complex because of the augmented degrees of freedom. In this
paper, we propose two heuristics to solve the resource allocation
problem that have very low computational complexity and give
performances not far from the optimal. The proposed heuristics
select a set of users for each subchannel, but contrary to the
reported methods that solve the throughput maximization problem,
our heuristics consider the set of real-time (RT) users to ensure
that their minimum rate requirements are met. We compare the
heuristics' performance against an upper bound and other methods
proposed in the literature and find that they give a somewhat
lower performance, but support a wider range of minimum rates
while reducing the computational complexity. The gap between the
objective achieved by the heuristics and the upper bound is not
large. In our experiments this gap is $10.7 \%$ averaging over all
performed numerical evaluations for all system configurations. The
increase in the range of the supported minimum rates when compared
with a method reported in the literature is $14.6\%$ on average.
\end{abstract}
\section {Introduction} \label{sec:Introduction}
With the ubiquitous use of smart phones, tablets, laptops and
Real-Time (RT) applications, traffic demand on the wireless access
network is increasing exponentially \cite{cisco-13}. In contrast,
mobile subscription prices have flattened in the last years
 due to competition and maturity of the market  \cite{wless-price}.
Therefore, there is a need to design systems that support high
data rates traffic with strict time deadlines, and concurrently
optimize the system resources to make deployments economically
profitable.

One of the key system design parameters in 4G wireless access
networks is spectral efficiency. Using spatial, user and frequency
diversity techniques in a multi-user Multiple Input Multiple
Output (MIMO)-OFDMA system, provides us with a high spectral
efficiency. These systems are proposed in current 4G standards,
such as Long Term Evolution (LTE) and IEEE 802.16~\cite{lte-rel10,
wim11-m}. However, when increasing the degrees of freedom for
transmission, a price has to be paid. Multi-antenna systems
require more hardware and software resources to process the
multiple spatial layers. In addition, the Resource Allocation (RA)
process becomes much more complex because we have many more
possibilities from which to choose.

The problem we deal with in this paper is the design of efficient
RA algorithms that provide us with solutions not too far from the
optimal, for a Zero-Forcing (ZF) Multiple Input Single Output
(MISO)-OFDMA system supporting minimum rates. This RA problem is a
nonlinear, non-convex integer program, which makes it almost
impossible to solve directly for any realistic number of
subchannels, users and antennas. For this reason, most research
work focuses on developing heuristic algorithms. It is also
important to benchmark the performance of these heuristic
algorithms. In \cite{perea13}, a dual method is proposed to find a
near-optimal solution to the sum rate maximization problem with
minimum rate constraints. It requires an enumeration of all
Spatial Division Multiple Access (SDMA) sets, which prevents the
method to be implemented efficiently, but it provides us an
off-line method useful for heuristic benchmarking.

Several heuristic methods have been used to solve the RA problem
for OFDMA-SDMA systems with both RT and non-Real Time (nRT)
traffic. In {\cite{tsai08}}, the objective is to maximize the sum
of the user rates subject to per-user minimum rate constraints
that model the priority assigned to each user at each frame. The
optimization problem is solved approximately for each frame by
minimizing a cost function representing the increase in power
needed when increasing the number of users or the modulation
order. The advantages of this approach are that it handles user
scheduling and RA together  and supports RT and nRT traffic. Its
weaknesses are that no comparison is made against a near-optimal
solution and the method used to determine user priorities at every
frame is very complex. In \cite{chung09}, both RT and nRT traffic
are supported. Priorities are set according to the remaining
deadline time for RT users and to the difference between the
achieved rate and the desired rate required for nRT users.
Comparisons against the algorithm in \cite{tsai08} show that the
packet drop rate for RT users and the algorithm's complexity are
significantly reduced. However, as in {\cite{tsai08}}, a
performance comparison with a near-optimal solution is not
provided.

In \cite{papoutsis10}, a heuristic algorithm is proposed for the
sum rate maximization problem with proportional rates among the
user data rates, i.e.,  the ratio among allocated user rates is
predetermined. The criteria used to form user groups includes
semi-orthogonality as in \cite{chung09}, but also fairness through
proportional rate constraints. This method is extended to include
hard minimum rates in \cite{papoutsis11}. There is no reported
method to evaluate the accuracy of these heuristics, except by
comparing them with each other.


In the heuristic method \cite{Lu2011}, the objective is the
weighted sum rate maximization under a total power constraint. The
user weights are updated at each frame to include different
fairness criteria. When compared to \cite{papoutsis10} the
performance is better, but Jain's fairness index is lower. The
sequential user selection to swap users require the channel matrix
inversion of all candidates, similarly to \cite{papoutsis11},
which creates a computational burden. In addition, for RT users
the method introduces delays by first detecting that the user
rates are lower than expected, and then adjusting the user
weights, i.e. no hard rate constraints are considered.

Our work differs from previously reported methods because we
consider hard minimum rate constraints for real-time users, which
is advantageous in terms of delay and QoS compliance. We propose
two efficient heuristic methods with much lower computational
complexity than the methods proposed in the literature. The
computational complexity reduction is several orders of magnitude
depending on the algorithm used and the problem parameters. We
compare the proposed heuristics performance against the
near-optimal solution proposed in \cite{perea13} and find that the
performance obtained is within $10.7 \%$ of the optimal averaging
over all performed numerical evaluations. In addition, the
proposed heuristics increase the range of the supported minimum
rates when compared with the method proposed in
\cite{papoutsis11}.  For the same case above, the increase in the
rate range is $14.6\%$ on average. This increase is achieved by
considering the rate constraint dual variables in the user power
allocation stage.

The paper is organized as follows, in section \ref{sec:prob-form},
we mathematically formulate the problem we want to solve: for a
given time slot, find the user selection and beamforming vectors
that maximize a linear utility function of the user rates, given a
total transmit power constraint and minimum rate constraints for
RT users. In section \ref{sec:heur}, we propose two heuristic
methods to solve the problem more efficiently. We compare their
performance against the upper bound and against other method
proposed in the literature in section \ref{sec:results}. Finally
in section \ref{sec:conclusions}, we summarize the main findings
and state our conclusions.
%
\section {Problem formulation and dual-based near-optimal method}
\label{sec:prob-form}
%
We consider the resource allocation problem for the downlink
transmission in a multi-carrier multi-user multiple input single
output (MISO) system with a single base station (BS). There are
$K$ users, some of which have RT traffic with minimum rate
requirements while the others have nRT traffic that can be served
on a best-effort basis. The BS is equipped with $M$ transmit
antennas and each user has one receive antenna. The system's
available bandwidth $W$ is divided into $N$ subchannels whose
coherence bandwidth is assumed larger than $W/N$, thus each
subchannel experiences flat fading. In the system under
consideration the BS transmits data in the downlink direction to
different users on each subchannel by performing linear
beamforming precoding. At each OFDM symbol, the BS changes the
beamforming vector for each user on each subchannel to maximize a
weighted sum rate. We assume that we use a channel coding that
reaches the channel capacity.

The BS transmits on each subchannel $n$, the signal vector
$\vect{x}_{n}=\sum_k\vect{w}_{n,k} s_{n,k}$, where $\vect{w}_{n,k}
\in \mathbb{C}^{M \times 1 }$ and $s_{n,k} \in \mathbb{C}$ are,
respectively, the beamforming vector and the information symbol
for user $k$ on subchannel $n$. The symbols $s_{n,k}$ are assumed
to be independent and follow the $\mathcal{CN}(0,1)$ distribution.
A power constraint $\sum_{n,k}\norm{ \vect{w}_{n,k} }^2 \leq
\check{P}$ is also imposed. The signal received at user $k$ on
subchannel $n$ is then given by
\begin{align}
y_{n,k}= \label{eq:y_k_rx} \vect{h}_{n,k} \vect{w}_{n,k} s_{n,k} +
\sum_{j \neq k} \vect{h}_{n,k} \vect{w}_{n,j} s_{n,j} + z_{n,k}.
\end{align}
where $\vect{h}_{n,k} \in \mathbb{C}^{1 \times M }$ is the channel
row $M$-vector between the BS and user $k$ on subchannel $n$, and
$z_{n,k}\sim\mathbb{CN}(0,1)$ is the white additive noise at the
receiver. The second term in Eq.~(\ref{eq:y_k_rx}) corresponds to
the inter-user interference. To simplify the RA problem, we assume
that the beamforming vectors are chosen according to the zero
forcing (ZF) criteria, which is known to be nearly optimal when
the SNR is high~\cite{yoo06}. For each subchannel, we can choose
at most $M$ users for which $\norm{ \vect{w}_{n,k} }^2
> 0$ and, for those users, the beamforming vectors must meet the
orthogonality constraints $\vect{h}_{n,k} \vect{w}_{n,j} = 0, \, j
\ne k$. Under the ZF constraint, the inter-user interference term
becomes zero in (\ref{eq:y_k_rx}) and the achievable rate of user
$k$ on subchannel $n$ is given by
\begin{equation}
\label{eq:bitrate} r_{n,k}\left(\vect{w_{n,k}}\right)= \log_2
\left( 1 + \norm{\vect{h}_{n,k} \vect{w}_{n,k}}^2 \right).
\end{equation}
%
The set of users $\mathcal{K}$ is divided into a set $\mathcal{D}$
of RT users with minimum rate constraints $\check{d}_{k}>0$ and a
set $(\mathcal{K}-\mathcal{D})$ of non real-time (nRT) users for
which $\check{d}_{k}=0$. The user selection is modelled by the
binary variables $\alpha_{k,n}$ which take the value $1$ when the
user $k$ is selected in the SDMA set of subchannel $n$, and zero
otherwise.

The objective of the RA algorithm is to maximize the weighted sum
rate of the users subject to the power, minimum rate and ZF
constraints. The users weights {$c_k$} and the minimum rate
constraints are determined by a higher layer scheduler.

%
Defining $\vect{w}, \boldsymbol{\alpha}$ as the vectors of stacked
optimization variables $\vect{w}_{n,k},\alpha_{n,k}$, the RA
problem can be mathematically formulated as follows:
\begin{align}
\label{eq:objzf} \max_{ \vect{w}, \boldsymbol{\alpha} }  \sum_{n=1,k=1}^{N,K} & c_k r_{n,k} ( \vect{w}_{n,k} ) \\
\sum_{n=1,k=1}^{N,K} \norm{ \vect{w}_{n,k} }^2 - \check{P}  \leq &
0
\label{eq:powerz} \\
- \sum_{n=1}^{N} r_{n,k} ( \vect{w}_{n,k} ) + \check{d}_{k} \leq &
0,
\quad k \in { \mathcal{D} } \label{eq:minratezf} \\
\sum_k \alpha_{n,k} & \le M,  \quad \forall n \label{eq:sdmasize}
\\
\iftoggle{onecolumnformat} {\left( \vect{h}_{n,k} \vect{w}_{n,j}
\right)^2 & \le B^{\prime} \left[ ( 1 - \alpha_{n,k}) +  ( 1 -
\alpha_{n,j} ) \right], \quad \forall n,\,\forall k,\, \forall j,
\, k \not= j \label{eq:zfconst} \\}
%
{\left( \vect{h}_{n,k} \vect{w}_{n,j} \right)^2 & \le B^{\prime}
\left[
( 1 - \alpha_{n,k}) +  ( 1 - \alpha_{n,j} ) \right],  \nonumber \\
& \quad \forall n,\, \forall k,\, \forall j, \, k \not= j
\label{eq:zfconst} \\}
\norm{\vect{w}_{n,k}} & \le A^{\prime} \alpha_{n,k} \label{eq:zfub} \\
\alpha_{n,k} & \in \left\{ 0, 1 \right\}  \label{eq:alphabin}
\end{align}
Constraint (\ref{eq:powerz}) is the total power constraint imposed
on the beamforming vectors and constraints (\ref{eq:minratezf})
assure that the RT users are assigned rates larger or equal than
their minimum rates $\check{\delta}_k$. Constraints
(\ref{eq:sdmasize}) to (\ref{eq:alphabin}) correspond to the ZF
constraints: Eq.~(\ref{eq:sdmasize}) guarantees that we do not
choose more than $M$ users for each subchannel,
Eq.~(\ref{eq:zfconst}) that two users in an SDMA set meet the ZF
constraints and Eq.~(\ref{eq:zfub}) that the beamforming vector is
null for users that are not in an SDMA set, $A^{\prime}$ and
$B^{\prime}$ are some large constants.

Problem~(\ref{eq:objzf}--\ref{eq:alphabin}) is a non-linear mixed
integer program (NLMIP). The vector of binary variables
$\boldsymbol{\alpha}$ determines the set of users that are
assigned to each subchannel. On the other hand, the vector of
continuous variables $\vect{w}$ determine the beamforming vectors
and need to comply with the ZF constraints
(\ref{eq:zfconst}--\ref{eq:zfub}) which depend on the user
selection binary variables $\boldsymbol{\alpha}$. There are many
off-the-shelf software packages available to solve NLMIPs, see
\cite{bussieck-12} for a survey. They use different methods with
different levels of accuracy and speed. However, the current NLMIP
solvers do not automatically exploit the specific structure of
problem~(\ref{eq:objzf}--\ref{eq:alphabin}). An off-line method is
proposed in \cite{perea13} to solve this problem with
near-optimality which we use to compare our heuristics.
\section{Efficient Heuristic methods}
\label{sec:heur}
In this section, we propose heuristic methods to solve
problem~(\ref{eq:objzf}--\ref{eq:alphabin}) efficiently.  We are
interested in feasible solutions, i.e., points that satisfy the
rate and power constraints, and that are not too far from the
optimal solution. In the dual-based near-optimal \cite{perea13},
power allocation and subchannel assignment are jointly performed.
Except for some trivial cases, we cannot separate the subchannel
allocation and power allocation processes. For heuristic methods,
however, we separate these processes in order to reduce
computational complexity. In the first stage, we find a subchannel
assignment that has enough subchannels assigned to the real-time
(RT) users, and in the second stage, we allocate power among users
using the fixed subchannel assignment. For the subchannel
assignment stage, we make use of the well known Semiorthogonal
User Selection (SUS) algorithm \cite{yoo06} to select user channel
vectors that have large norms and are semiorthogonal to each
other. But contrary to the throughput maximization case, we
include the RT users to satisfy their minimum rates when selecting
the user set for each subchannel. For the power allocation stage,
we use a method that finds feasible points and is much quicker
than optimally solving the power optimization problem. The
subchannel assignment algorithm and the power allocation algorithm
constitute the proposed heuristic method.

\subsection{General Description of the Proposed Heuristic Method}
The basis for the design of our heuristic is the realization that
for a fixed subchannel assignment $\boldsymbol{\alpha}$ in
problem~(\ref{eq:objzf}--\ref{eq:alphabin}), the resulting power
allocation problem can be approximated to a convex one which is
much easier to solve. Thus, instead of enumerating all feasible
values of variable $\boldsymbol{\alpha}$ corresponding to every
subchannel assignment, our heuristic chooses a subchannel
assignment and solves a convex power allocation problem. Then, it
reassign resources and solves the power allocation problem again
until the RT users' rate constraints are met.

There are two mechanisms to reassign resources to users. The first
mechanism --- \emph{subchannel reassignment} --- takes away
subchannels assigned to users that do not require them, because
they are not RT users or they have more resources than needed, and
assigns them to the users in need. The second mechanism ---
\emph{rate-constrained power allocation} --- takes into account
the user rate constraints to reallocate power between users.
Subchannel reassignment has a much larger effect because users in
need are given subchannels that they did not have before; the
rates increase substantially with every subchannel added.
Rate-constrained power allocation has a lower effect because the
rate increase dependency against power is logarithmic. However,
this mechanism proves to be crucial in finding feasible points
when the minimum rate requirements increase. In addition,
recomputing the users power is quicker than finding a new
subchannel and inverting its new channel matrix.

\begin{figure}
\centering
\includegraphics[scale=0.5]{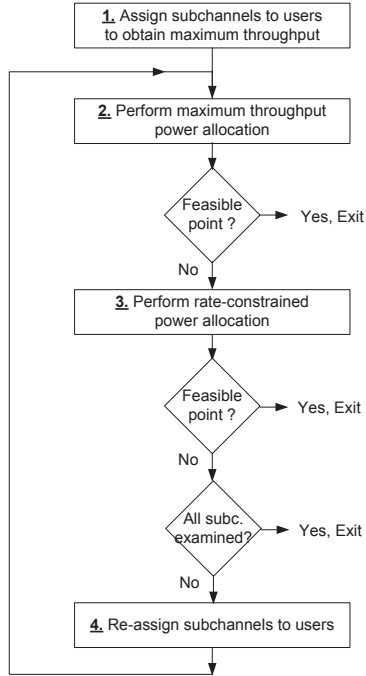}
\caption [Heuristic general algorithm] {Heuristic general
algorithm} \label{fig:heur-diag}
\end{figure}

The proposed heuristic method starts by solving
problem~(\ref{eq:objzf}--\ref{eq:alphabin}) without considering
rate constraints (\ref{eq:minratezf}). If the required rates
$\check{d}_k$  are lower or equal than the obtained rates, we have
an optimal solution and the algorithm finishes. To obtain the
maximum throughput solution efficiently, we use the SUS algorithm
to assign subchannels to users and then perform \emph{maximum
throughput power allocation}, which consists of finding the user
power allocation that satisfies the power constraint with equality
disregarding the rate constraints. These correspond to the first
two blocks in the diagram of figure \ref{fig:heur-diag}. If the
required rates are met, we exit, otherwise, we need to assign more
resources to the users in need, thus we perform rate-constrained
power allocation as indicated by block 3 in figure
\ref{fig:heur-diag}.

We perform subchannel \emph{re}-assignment when the maximum
throughput subchannel assignment plus rate-constrained power
allocation does not support the required minimum rates. A
heuristic method that groups semiorthogonal user vectors is used
to assign more subchannels to the users in need and it is
indicated by block 4 in figure \ref{fig:heur-diag}. We perform
iterations adding subchannels to users in need and performing
power allocation (blocks 2 and 3) until the user minimum rates are
met or there are no more subchannels to reassign and the problem
is declared unfeasible by the heuristic.
\subsection{Power Allocation}
\label{subsec:pow-all} In this subsection we find the user power
allocation for a fixed subchannel assignment. Assume that we have
chosen a vector $\boldsymbol{\alpha}$  for each subchannel $n$
satisfying Eqs.~(\ref{eq:sdmasize}) and~(\ref{eq:alphabin}. We
explain the heuristic method to obtain such a vector in section
\ref{sec:subc-ass}. The vector $\boldsymbol{ \alpha}^{(n)}$
determines a fixed SDMA set of users, $S_n$, defined as
\begin{align}
\label{eq:sdma-def} S_n &\doteq \left\{  k \in \mathcal{K} :
\alpha_{n,k}= 1 \right\}, \\ g_n &\doteq |S_n|, \quad \forall n.
\end{align}
Sets $S_n$ contain the indexes of the users assigned to subchannel
$n$. We first reformulate problem
(\ref{eq:objzf}--\ref{eq:alphabin}) using these known sets. Then,
we apply a dual method to solve it. For this purpose, we arrange
the channel vectors
 of selected users in the rows of a $g_n \times M$ matrix
\begin{align}
\label{eq:H-def} \vect{H}_n \doteq \begin{bmatrix} \vect{h}_{ n,
S_n(1) }
\\
\vdots \\ \vect{h}_{ n, S_n(g_n) } \end{bmatrix}, \quad \forall n,
\end{align}
where $S_n(j)$ is the $j$-th user in the set $S_n$.  We also
arrange the corresponding beamforming vectors in the columns of a
$M \times g_n$ matrix for each subchannel
\begin{equation}
\label{eq:W-def} \vect{W}_n \doteq \left[ \vect{w}_{n,S_n(1)},
\ldots, \vect{w}_{n,S_n(g_n)} \right], \quad \forall n.
\end{equation}
Then, the ZF constraints (\ref{eq:zfconst}) can be written as
\begin{align}
\label{eq:ZF-reform} \vect{H}_n \vect{W}_n  = \diag ( \sqrt{
\vect{q}^{(n)} } ), \quad \forall n
\end{align}
where $\vect{q}^{(n)} = \{q_{n,j}\}$ is the users power vector
comprised of
\begin{align}
q_{n,j} = \vect{h}_{n, s_n(j)} \vect{w}_{n, s_n(j)}, \quad j \in
\{ 1, \ldots, g_n \}
\end{align}
Beamforming vectors for users $k$ not belonging to  $s_n$ are set
to zero. Restricting the direction of $\vect{W}_n$ to the
pseudo-inverse of matrix $\vect{H}_n$ as done in \cite{perea13},
we obtain from (\ref{eq:ZF-reform})
\begin{align}
\label{eq:W-direct} \vect{W}_n = \vect{H}_n^{\dag}  \diag ( \sqrt{
\vect{q}^{(n)} } ), \quad \forall n
\end{align}
the power constraint can now be written as
\begin{align}
\label{eq:pow-const-reform} \sum_{n=1}^N \tr{( \vect{W}_n^H
\vect{W}_n )} - \check{P} \leq 0
\end{align}
and replacing (\ref{eq:W-direct}) in (\ref{eq:pow-const-reform})
we obtain
\begin{align}
\label{eq:pow-const-reform2} \sum_{n=1}^N \sum_{j=1}^{g_n} {\left[
(\vect{H}_n^{\dag})^{H} \vect{H}_n^{\dag} \right]}_{j,j} q_{n,j} -
\check{P} \leq 0
\end{align}
Let's define the entries of the $N \times K$ matrices
$\boldsymbol{\beta}$ and $\vect{p}$ as
\begin{align}
\label{eq:mat_gamma_def}
 \beta_{n,k} \doteq &
\begin{cases}
{\left[ (\vect{H}_n^{\dag})^{H} \vect{H}_n^{\dag} \right]}_{j,j}
\quad \text{if } k= s_n(j), \quad \forall j \in \{1,\ldots,g_n\}
\\
0,  \quad \quad \quad \quad  \quad \quad \text{otherwise}
\end{cases}
\\ \nonumber
\\
\label{eq:matp_def} p_{n,k} \doteq &
\begin{cases} q_{n,j} \quad \text{if } k= s_n(j), \quad \forall  j \in \{1,\ldots,g_n\}
\\
0, \quad \quad \text{otherwise.}
\end{cases}
\end{align}
The power constraint can now be expressed as
\begin{align}
\label{eq:pow-const-reform3} \sum_{n=1}^N \sum_{k=1}^{K}
\beta_{n,k} p_{n,k} - \check{P} \leq 0.
\end{align}
From the original model (\ref{eq:objzf}--\ref{eq:alphabin}), we do
not need constraints (\ref{eq:sdmasize}) and (\ref{eq:alphabin})
because we choose $\alpha_{n,k}$ satisfying these conditions.
Constraints (\ref{eq:zfconst}) and (\ref{eq:zfub}) are implicit in
the reformulated model because the beamforming vectors satisfy
(\ref{eq:ZF-reform}) and we set to zero all beamforming vectors
for which $\alpha_{n,k}=0$. Therefore, only constraints
(\ref{eq:powerz}) and (\ref{eq:minratezf}) remain. We also changed
the problem optimization variables from the vectors
$\vect{w}_{n,k}$ to the scalars $p_{n,k}$ because the vector
directions are now fixed by (\ref{eq:W-direct}).

Replacing $(\vect{h}_{n,k} \vect{w}_{n,k})^2$ by $p_{k,n}$
in~(\ref{eq:objzf}),(\ref{eq:minratezf}) and
replacing~(\ref{eq:powerz}) by~(\ref{eq:pow-const-reform3}), we
obtain the problem formulation
\begin{align}
  \label{eq:sub-problem-q}
  \max_{ p_{n,k} }  \quad \sum_{n=1}^N \sum_{k=1}^K  c_k
  & \log_2( 1 + p_{n,k} )  \\
  \sum_{n=1}^N \sum_{k=1}^K  \beta_{n,k} \,\ p_{n,k}
  -  \check{P} & \leq 0 \label{eq:subp-q-pow} \\
 -\sum_{n=1}^N \log_2( 1 + p_{n,k} ) + \check{d}_k & \leq 0,
\quad k \in { \mathcal{D} } \label{eq:subp-q-rat} \\
p_{n,k} & \geq 0, \quad \forall n,k.  \label{eq:subp-box}
\end{align}

\subsection{Optimal Power Allocation} \label{subsec:opt-power-all}
Problem~(\ref{eq:sub-problem-q}--\ref{eq:subp-box}) is convex
since it maximizes a concave function over a convex set formed by
constraints (\ref{eq:subp-q-pow}--\ref{eq:subp-box}). We can solve
this problem optimally using a dual Lagrange approach. First, we
define dual variables $\theta$ for the power constraint
(\ref{eq:subp-q-pow}), and $\{ \delta_k \}$ for the rate
constraints (\ref{eq:subp-q-rat}). Then, we derive a closed-form
expression of the dual function and solve the dual problem. This
yields the water-filling power allocation \cite{perea-pow2013}
\begin{align}
\label{eq:optp-def1} \overline{p}_{n,k} =& { \left[ \frac{ c_k+
\delta_k } { \theta \beta_{n,k} \ln2 } -1 \right] }^+, \quad
\forall n, \forall k: \beta_{n,k} \neq 0
\end{align}
and rate allocation
\begin{align}
\label{eq:rate-computation} r_{n,k} =& \log_2 \left( 1 + { \left[
\frac{ c_k+ \delta_k } { \theta \beta_{n,k} \ln2 } -1 \right] }^+
\right), \quad \forall n,
\end{align}
where $\boldsymbol{\delta}$ is the vector of dual variables
$\delta_k$. For convenience we have defined dual variables
$\delta_k$ for all users including the ones with no minimum rate
requirements ($k \notin { \mathcal{D} }$); for these users we set
$\check{d}_k=0$.

We can the optimal dual variables $\theta,\{ \delta_k \}$ using
derivative-free techniques. The use of such methods involves two
steps. In the first one, we compute a matrix pseudo-inverse per
subchannel to obtain the inverse of the channel effective gains
$\beta_{n,k}$; this has computational complexity $O(NM^3)$. In the
second step, we perform subgradient iterations computing the power
and rate constraints to obtain the subgradient vector; this has
computational complexity is $O(N(M+D)I_d)$, assuming a maximum
number of subgradient iterations $I_d$.
\subsection{Efficient Power Allocation} \label{sec:heur-pow-all}
To solve problem~(\ref{eq:sub-problem-q}--\ref{eq:subp-box}) more
efficiently we separate it in two stages: maximum-throughput power
allocation (PA) and rate-constrained PA. Maximum-throughput PA
only considers the power constraint (\ref{eq:subp-q-pow}), thus
$\delta_k=0$ in (\ref{eq:optp-def1}) and we just need to find the
dual variable $\theta$ that satisfies the power constraint
 with equality. For this purpose, we use the
exact method reported in \cite{palomar-05, perea-pow13} which is
summarized below
\begin{enumerate}
\item Find the dual variable $\theta$ that satisfies the power
constraint (\ref{eq:subp-q-pow}) with equality, using
\begin{equation}
\label{eq:theta-maxthr-heur-comp} \theta^{(i)} = \frac{
\sum_{k=1}^K | \mathcal{B}_k( \check{\theta}, \delta_k ) | c_k } {
( \check{P} + \sum_{k=1}^K \sum_{n\in\mathcal{B}_k(
\check{\theta}, \delta_k ) } \beta_{n,k} )\ln 2 },
\end{equation}
where $\check{\theta}>0$ is a lower bound of $\theta$,
$\delta_k=0$, $i$ is the iteration index and
\begin{eqnarray}
\label{eq:setC-def-1} \mathcal{B}_k( \theta, \delta_k )  \doteq
\big\{ n \in \mathcal{N}  : ( k \in S_n ) \land \\
\nonumber ( \beta_{n,k} < \frac{(c_k+\delta_k)}{ \theta \ln2} )
\big\}, \quad \forall k \in \mathcal{K},
\end{eqnarray}
with $S_n$ is the SDMA set associated to subchannel $n$, and
$\mathcal{N}$ the set of all subchannels.
\item Recompute sets $\mathcal{B}_k^{(i+1)}$ using $\theta^{(i)}$.
If the sets $\mathcal{B}_k^{(i)}$ and $\mathcal{B}_k^{(i+1)}$ are
equal, we have found the solution, the power constraint is
satisfied with equality. Otherwise, iterate recomputing
(\ref{eq:theta-maxthr-heur-comp}) and (\ref{eq:setC-def-1}) until
 finding identical sets in two consecutive iterations.
\end{enumerate}
After solving the maximum throughput PA problem, if the achieved
rates are feasible they correspond to the optimal ones. If the
rate constraints are not met, we incorporate the rate
constraints~(\ref{eq:subp-q-rat}) and perform rate-constrained PA
using the heuristic reported in \cite{perea-pow13}. This method
finds a feasible point and does not require any iteration. This
feasible point satisfies the power constraint with equality but
the rate constraints with inequality and is faster to compute than
the subgradient algorithm. Here, we summarize the method
\begin{enumerate}
\item Compute the set of unsatisfied users $\mathcal{T}$ as
\begin{eqnarray}
\label{eq:rate-feas-cond} \mathcal{T} \doteq \{ k \in
\{1,\ldots,K\} : r_k < \check{d}_k \} \\
\label{eq:rate-feas-def}r_k \doteq \sum_{n=1}^N \log_2 \left( 1 +
{\left[ \frac {c_k} {\theta^{(1)} \beta_{n,k} \ln2} - 1 \right]}^+
\right)
\end{eqnarray}
where $\theta^{(1)}$ is the optimal power allocation dual variable
obtained after maximum throughput PA.
\item For the users that do not belong to $\mathcal{T}$, make
$\delta_k=0$. For the other users, obtain the minimum value of the
dual variable $\delta_k$ required to satisfy the rate constraints
\begin{equation}
\label{eq:delta_heur_comp} \delta_k^{(2)}= {\left[ ( \bar{\theta}
\ln2 ) \left( 2^{\check{d}_k} \prod_{n \in \mathcal{A}_k^{\prime}}
\beta_{k,n} \right)^{{| \mathcal{A}_k^{\prime} |}^{-1}} - c_k
\right]}^+
\end{equation}
where $\mathcal{A}_k^{\prime}= \mathcal{B}_k(\bar{\theta},
\delta_k)$, $\bar{\theta}$ is an upper bound of $\theta$ given by
\begin{equation}
\label{eq:bar-epsilon} \bar{\theta} = \theta^{(1)}
2^{(\check{d}_k- r_k)\epsilon}
\end{equation}
and $\epsilon>0$ is a parameter found experimentally as explained
in subsection \ref{subsec:PA-perf}.
\item Using dual variables $\delta_k^{(2)}$, compute the power
constraint dual variable $\theta$ that satisfies the constraint
(\ref{eq:subp-q-pow}) with equality.
\end{enumerate}
Maximum throughput and rate constrained power allocation
correspond to blocks 2 and 3 in figure \ref{fig:heur-diag}.
%
\subsection{Subchannel Assignment Heuristic} \label{sec:subc-ass}
The purpose of block 1 in figure \ref{fig:heur-diag} is to perform
user subchannel assignment to obtain high rates. The rates in
(\ref{eq:rate-computation}) are affected by the effective channel
gains $\beta_{n,k}^{-1}$, which increase when the channel vector
norms are large and the chosen vectors for each subchannel are
semi-orthogonal to each other. We use the well-known SUS algorithm
\cite{yoo06} to perform this assignment. The computing efficiency
of this algorithm is improved in \cite{mao12}. We rewrite the SUS
algorithm splitting it in two parts: an initialization stage and a
user search stage. This is done to adapt the SUS algorithm to the
rate-constrained case described in subsection
\ref{subsec:proposed-subc-reass}.

The input to SUS initialization stage is the set of available
users $\mathcal{U}_0$. Its output are the user with the maximum
norm $\{\pi_0\}$ and a matrix $\vect{G}_0$ forming a basis of the
null space spanned by the channel vector $\vect{h}_{\pi_0}$. We
write the input/output relation of this stage as [ $S_n^0,
\vect{G}_n^0$ ]= SUS\_init $( \mathcal{U}_0 )$, where $S_n^0$
contains only the selected first user $\{\pi_0\}$.

In the SUS search stage, we compute the projection of the
remaining channel vectors to the null space spanned by the channel
vectors of the users already selected. We pick the user whose
projection is the largest, add it to set $S_n$ and recompute
matrix $\vect{G}_n$. We add users until the set contains $M$ users
or we finish examining all users in the input set $\mathcal{U}_0$.
The input/output relation of the users search stage is [$S_n,
\vect{G}_n$]= SUS\_search ( $\mathcal{U}_0$, $S_n^0$,
$\vect{G}_n^0)$, where $S_n$ is the set of selected users for
subchannel $n$ and $\vect{G}_n$ its the matrix that spans the null
space of the selected user vectors. The SUS algorithm as described
in \cite{yoo06} is implemented by making
$\mathcal{U}_0=\{1,\ldots,K\}$ and sequentially invoking the two
stages: [$S_n^0$, $\vect{G}_n^0$]= SUS\_init ($\mathcal{U}_0$) and
[$S_n, \vect{G}_n$]= SUS\_search ( $\mathcal{U}_0-S_n^0$, $S_n^0$,
$\vect{G}_n^0$).

After performing subchannel assignment for each subchannel $n$, we
perform maximum throughput and rate constrained power allocation
--- indicated by blocks 2 and 3 in figure \ref{fig:heur-diag} ---
using the heuristics described in \ref{sec:heur-pow-all}.

\subsection{Subchannel Reassignment Heuristic}
\label{subsec:proposed-subc-reass}
%

\begin{algorithm}
\begin{algorithmic}
\STATE { \underline{Input:} Current dual variables
$\theta,\{\delta_k\}$, current subchannel assignment sets $S_n$}
\STATE { \underline{Output:} New subchannel assignment sets $S_n$,
user rates $r_k^{(1)}$ or $r_k^{(2)}$}
\STATE { - Compute $r_{n,k}$ and set of users in need
$\mathcal{T}$ using (\ref{eq:rate-feas-cond})}
\STATE { - Order subchannels according to maximum norm of the
users in need, i.e. according to $ \max_{k\in{\mathcal{T}}} \norm
{ h_{n,k}}, \, \forall n$, producing ordered set $\mathcal{N}$ }
\FORALL { $n \in \mathcal{N}$ }
        \STATE {- Compute critical user set $\mathcal{E}$ containing users in $S_n $ for which $ \sum_{m \neq n} r_{m,k} < \check{d}_k$ }
        \STATE {If $\mathcal{E} = \emptyset$, $\mathcal{Z} \leftarrow \mathcal{T}$  }
        \STATE {Otherwise, $\mathcal{Z} \leftarrow \mathcal{E}$  }

        \STATE { [$S_n^0, \vect{G}_n^0$]= SUS\_Init
        $(\mathcal{Z})$ }
        \STATE { [$S_n, \vect{G}_n$]= SUS\_search ( $\mathcal{Z}-S_n^0, S_n^0, \vect{G}_n^0)
        $}
        \IF { $|S_n| < M$ }
            \STATE { [$S_n, \vect{G}_n$]= SUS\_search $(\{1,\ldots,K\}-S_n,S_n,\vect{G}
            )$ }
        \ENDIF

        \STATE { Compute pseudo-inverse of channel matrix formed by users
in $S_n$ and compute $\beta_{n,k}$ using (\ref{eq:mat_gamma_def})
}

        \STATE { Perform Maximum Throughput power allocation obtaining
                dual variable $\theta^{(0)}$ that satisfies power constraint with
                equality, obtaining rates  $r_k^{(1)}$ }
        \IF { $r_k^{(1)}$ satisfy rate constraints }
                \STATE{ Exit}
        \ELSE
                \STATE { Perform rate constrained power allocation, obtaining rates $r_k^{(2)}$ }
                \IF { $r_k^{(2)}$ satisfy rate constraints }
                    \STATE{ Exit}
                \ENDIF
        \ENDIF

        \STATE { Update set of users in need, $\mathcal{T}$ }

\ENDFOR
\end{algorithmic}
\caption{Subchannel Reassignment Heuristic Algorithm}
\label{algo:heur-subc-reass}
\end{algorithm}
%

After executing block 3 of the diagram in figure
\ref{fig:heur-diag}, we obtain the user rates per subchannel
$r_{n,k}^{(0)}$. The user rates are simply computed by
$r_{k}^{(0)}=\sum_{n \in \mathcal{C}_k} r_{n,k}^{(0)}$, where
$\mathcal{C}_k$ is the set of subchannels assigned to user $k$. We
compute the set of unsatisfied users $\mathcal{T}$ given by
(\ref{eq:rate-feas-cond}). Set $\mathcal{T}$ tells us which users
need to be assigned additional subchannels. We first scan the
subchannels in which any of the users in need have good channel
conditions, so they can be first reassigned to these users. The
computational complexity of the subchannel ordering is bounded by
$O(N^2)$. For each subchannel $n$, we build a critical set
$\mathcal{E}$ containing the users in the current SDMA set that
can not be removed from the SDMA set because that would take the
user out of feasibility. We consider two cases: first that the set
$\mathcal{E}$ is empty. In this case we invoke the SUS
initialization algorithm to select the strongest user in
$\mathcal{T}$ as the first element of the SDMA set. Then, we
invoke the SUS search stage to add users to the SDMA set. We
initially scan other users in $\mathcal{T}$ so that they can be
added with priority to the SDMA set, and if the SDMA set has not
yet been completed, we scan the remaining users $\{1,\ldots,K\} -
\mathcal{T}$.

In the second case, when set $\mathcal{E}$ is not empty, we
initialize the SDMA set with all the users in set $\mathcal{E}$,
and then add users invoking the SUS search algorithm. To add users
to this SDMA set, we scan the rest of the users but look first in
the set of users in need $\mathcal{T}$. The difference between the
cases $\mathcal{E} = \emptyset$ and $\mathcal{E \neq \emptyset}$
is that in the second case, we keep the users in need that are
already in the SDMA set before trying to add more users.

Notice that all the selected users comply with the
semiorthogonality condition of the SUS algorithm; the only change
in computations to the SUS algorithm is the order in which we
examine the users. By changing the order, we are giving priority
to the users in need. The computational complexity of this stage
is bounded by the complexity of the maximum throughput SUS search
algorithm, i.e. $O(KM^3)$.

After obtaining the new SDMA set for a subchannel, we perform
maximum throughput power allocation. If the resulting rates are
feasible we exit the algorithm. Otherwise, we perform
rate-constrained power allocation. If the resulting rates are
still not feasible, we continue reassigning subchannels until the
rates are feasible or there are no more subchannels and the
algorithm declares that is not able to find a feasible point. This
corresponds to the loop in the lower part of the block diagram in
figure \ref{fig:heur-diag}. The pseudo-code of the subchannel
reassignment heuristic is listed in algorithm
\ref{algo:heur-subc-reass}, which corresponds to the sequence of
blocks 4,2,3 in figure \ref{fig:heur-diag}.

The power allocation  algorithms have computational complexity
$O(KN)$ \cite{perea-pow13} and the SUS search algorithm has
complexity $O(KM^3)$. Assuming the worst case where all
subchannels are examined for reassignment, the proposed
algorithm's overall computational complexity is
%
\begin{eqnarray}
\label{eq:reass-complexity} O_{\text{alg.
}\ref{algo:heur-subc-reass}}=
\begin{cases}
O(KN^2), & \text{if } N > M^3
\\
O(KNM^3), & \text{otherwise.}
\end{cases}
\end{eqnarray}
This is lower than Papoutsis' method \cite{papoutsis11}
computational complexity, which is $O(KN^2M^4)$ for all $N$.
\begin{table}
\centering
\begin{tabular}{| >{\centering\arraybackslash}m{1in} | >{\centering\arraybackslash}m{0.8in} |>{\centering\arraybackslash}m{1.2in} |}

 \hline
Algorithm  & Complexity & Purpose \\
 \hline\hline
Alg. \ref{algo:heur-subc-reass} & Eq. (\ref{eq:reass-complexity})
& Prob.~(\ref{eq:objzf}--\ref{eq:alphabin})\\ \hline Alg. 2 &
$O(KNM^3)$ & Prob.~(\ref{eq:objzf}--\ref{eq:alphabin}) with
per-subchannel power constraint \\ \hline Papoutsis' method
\cite{papoutsis11} & $O(KN^2M^4)$ &
Prob.~(\ref{eq:objzf}--\ref{eq:alphabin}) with per-subchannel
power constraint \\ \hline Dual bound \cite{perea13} & $O(NK^M
M^3)$ & Dual of (\ref{eq:objzf}--\ref{eq:alphabin}) \\ \hline
 \hline
 \end{tabular}
\caption  {Algorithms complexity}
 \label{tab:myalgs-comp}
\end{table}
\subsection{ Reduced Complexity Algorithm}
\label{subsec:per-subc-PA} In this section, we devise a variation
to the subchannel reassignment algorithm
\ref{algo:heur-subc-reass} that linearizes the dependency of the
computational complexity in expression (\ref{eq:reass-complexity})
with respect to the number of subchannels $N$, for $N > M^3$.
Since in LTE-Advanced systems, the maximum number of subchannels
is large, it is important to linearize the computational
complexity with respect to $N$.

For this purpose, we solve a sum rate maximization problem with
one power constraint per subchannel instead of a total power
constraint and we do not consider the rate
constraints~(\ref{eq:subp-q-rat}). In the subchannel iteration
loop in algorithm \ref{algo:heur-subc-reass}, we update the user
power corresponding to all subchannels because the
power-constraint dual variable affects them all, which produces
the term $N^2$ in the complexity expression
(\ref{eq:reass-complexity}) for $N > M^3 $. To solve the new
problem formulation, we need to update the user power
corresponding to only one subchannel. Therefore, when computing
the power and rates after power allocation, only the ones
corresponding to that subchannel are affected, making the
computational complexity of this step $O(K)$ as opposed to the
original $O(KN)$. The computational complexity of this method is
\begin{equation}
\label{eq:nodual-complexity} O_{\text{alg. 2}} = O(KNM^3).
\end{equation}
(\ref{eq:nodual-complexity}) varies linearly with $N$ and since
$N$ ranges from $6$ to $550$ in a LTE-Advanced system with Carrier
Aggregation (CA) \cite{lte-rel10}, this results in a much faster
algorithm for large $N$. The reduced complexity method does not
support the high minimum rates that algorithm
\ref{algo:heur-subc-reass} can, but it is a more efficient
algorithm when $N$ is large. We name this simplified method
algorithm 2, but we do not provide its pseudo-code since the
differences with algorithm \ref{algo:heur-subc-reass} are
straightforward. Table \ref{tab:myalgs-comp} summarizes the
computational complexity of the methods presented in this paper.
\section{Numerical Evaluations}
\label{sec:results} In this paper we devised heuristic algorithms
to efficiently solve problem~(\ref{eq:objzf}--\ref{eq:alphabin}).
For this purpose, we proposed algorithm \ref{algo:heur-subc-reass}
in subsection \ref{subsec:proposed-subc-reass} and its simplified
version, algorithm 2, in subsection \ref{subsec:per-subc-PA}.
Table \ref{tab:myalgs-comp} shows that they have reduced
computational complexity when compared to other methods. In this
section, we numerically evaluate their performance and CPU load,
and compare them against existing methods. The power allocation
(PA) heuristics presented in subsection \ref{sec:heur-pow-all} are
invoked at each subchannel assignment iteration as illustrated in
the block diagram of figure \ref{fig:heur-diag}. Thus, the CPU
load and performance of the PA algorithms have an important effect
on the overall heuristic performance. In subsection
\ref{subsec:PA-perf} we evaluate the performance of these PA
heuristics independently of the subchannel assignment method, we
assume certain subchannel assignment and evaluate the PA
heuristics.

In subsection \ref{subsec:methods-perf} we evaluate the
performance of the proposed overall heuristics and the support of
the fulfilled minimum rates. Our interest is on answering the
following questions: how far from the optimal is the sum rate
achieved by these methods; how fast are the proposed heuristics
compared to existing methods; what is the range of the minimum
rates supported by the rate-constrained PA methods, as opposed to
the maximum throughput PA methods; and how the overall heuristics
compare to \cite{papoutsis11}.

\subsection{ Performance of Power Allocation Heuristics} \label{subsec:PA-perf}
\begin{figure}
\centering
\includegraphics[scale=0.33]{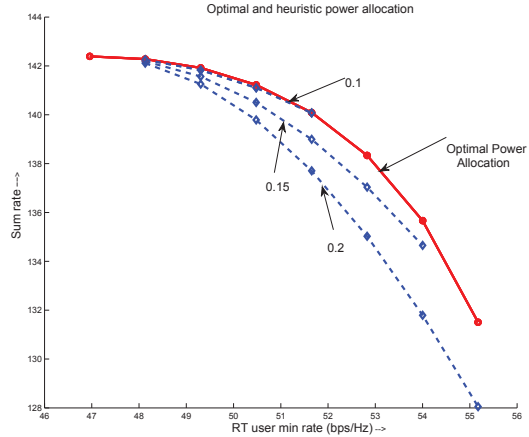}
\caption [Optimal and heuristic power allocation comparison for
different values of $\epsilon$]  {Optimal and heuristic power
allocation comparison for different values of $\epsilon$. $K=3,
M=3, N=1, \check{P}=20$.} \label{fig:pow-all-varE}
\end{figure}

%
%
Figure \ref{fig:pow-all-varE} shows an example of the performance
obtained by the PA heuristics presented in subsection
\ref{sec:heur-pow-all} for the listed system parameters. In the
figure, we compare the sum rate given by the optimal solution to
problem~(\ref{eq:sub-problem-q}--\ref{eq:subp-box}) and the result
given by the PA heuristics. For minimum rate constraints lower
than $47$ bps/Hz, the rate constraints are inactive and the
maximum throughput PA heuristic method gives the optimal sum rate.
However, for minimum rate constraints higher than $47$ bps/Hz, the
performance of the rate-constrained PA heuristic is sub-optimal
and it depends on the parameter $\epsilon$ in
(\ref{eq:bar-epsilon}). From the results of figure
\ref{fig:pow-all-varE}, we observe that a parameter value close to
$\epsilon=0.2$ achieves a large minimum rate support. For the
example in figure \ref{fig:pow-all-varE}, rate-constrained PA
extends the range of the minimum rates constraints from $47$ to 56
bps/Hz, i.e.,$16$\%. Evaluating the performance over multiple
configurations, we find that rate-constrained PA increases the
range of supported minimum rates between $15$\% and $30$\% when
compared to schemes that perform maximum throughput PA only. In
general, any subchannel assignment method can increase the
supported minimum rates by performing rate-constrained PA as we
show below.
%
\subsubsection{Benefits of rate-constrained power allocation}
\begin{figure}
\centering
\includegraphics[scale=0.525]{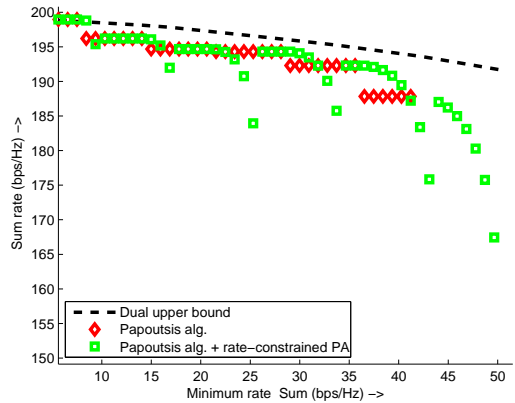}
\caption [Optimal and heuristic methods comparison] {Optimal and
heuristic methods comparison. $K=16, M=3, N=8, D=1, \check{P}=20$.
} \label{fig:papoutsis-dual-ext}
\end{figure}

Papoutsis method \cite{papoutsis11} only performs maximum
throughput PA. If we add rate-constraints PA to Papoutsis method,
we would achieve and increase in the range of minimum rate
constraints supported. This is illustrated in figure
\ref{fig:papoutsis-dual-ext} for the system parameters listed,
where the minimum rates support is extended from $40$ bps/Hz to
$50$ bps/Hz, i.e., a $20$\% increase. In this example, we
optimally solve a power allocation problem similar
to~(\ref{eq:sub-problem-q}--\ref{eq:subp-box}), but with
sub-channel power constraints instead of a total power constraint.
The smearing effect in figure \ref{fig:papoutsis-dual-ext} occurs
when the rate constraints are not satisfied with simple maximum
throughput PA and rate-constrained PA is applied.

It is not possible, however, to solve the rate-constrained PA
problem without an increase in the computational complexity. For
this reason, we use the PA heuristics described in section
\ref{sec:heur-pow-all}.
\subsection{ Overall Heuristics Performance Comparison} \label{subsec:methods-perf}
We use a Rayleigh fading channel model to generate independent
channels and compare numerically the supported minimum rates and
the sum rate achieved
 by the following methods:
\begin{enumerate}
\item Dual-based upper bound \cite{perea13}
\item Papoutsis' algorithm \cite{papoutsis11}
\item The proposed heuristic SUS-based heuristic algorithm
\ref{algo:heur-subc-reass} described in subsection
\ref{subsec:proposed-subc-reass}
\item A simplified SUS-based heuristic algorithm 2, described in
subsection \ref{subsec:per-subc-PA} that performs PA considering
per-subchannel power constraints.
\end{enumerate}

\begin{figure}
\centering
\includegraphics[scale=0.6]{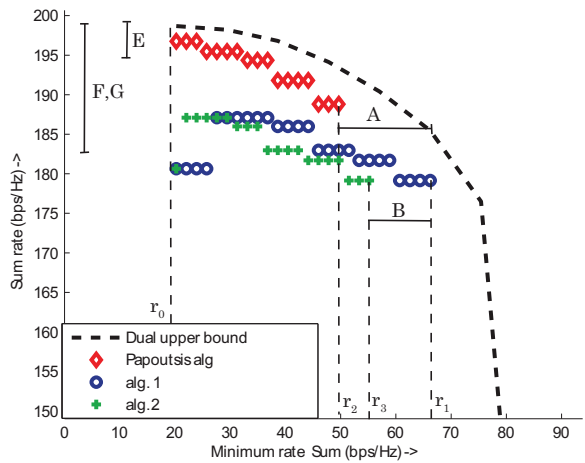}
\caption [Optimal and heuristic methods comparison] {Optimal and
heuristic methods comparison.} \label{fig:subc-ass-meth-comp}
\end{figure}

Figure \ref{fig:subc-ass-meth-comp} illustrates one example of the
objective achieved by these methods for one real-time user, $D=1$
and parameters $N=8$, $K=8$, $M=3$, $\check{P}=20$. The plots only
show feasible points. Thus, when increasing the minimum required
rates (horizontal axis), the curves stop if the methods can no
longer find feasible points.

We start by solving the problem without considering minimum rate
requirements as indicated by blocks 1 and 2 in figure
\ref{fig:heur-diag}; this gives us user rates $\{r_k^0\}$. If we
were to extend the curves of figure \ref{fig:subc-ass-meth-comp}
to zero, they would be flat curves with $\sum_k r_k^0$ as the sum
rate. We want to focus on the domain where rate constraints are
active. For this purpose, we increase the rate constraints by
increments $\Delta_r$, i.e., $\check{d}_k= r_k^0 + \Delta_r$, for
the RT users and solve the problem for each $\check{d}_k$.

For a number of RT users $D>1$, we use $\sum_{k=1}^D \check{d}_k$
to list the minimum rate constraint in plots. In our numerical
evaluations, we increase the rate constraints and try to find
feasible points using the heuristics until the dual upper bound
becomes positive indicating the problem unfeasibility. In figure
\ref{fig:subc-ass-meth-comp}, the upper bound provided by the
negative of the dual function maximum is shown by a dashed line
and it is the reference to measure the performance of all
heuristic methods. Papoutsis' method is shown in diamond markers
and closely follows the upper bound. The proposed SUS-based
heuristic (algorithm \ref{algo:heur-subc-reass}) is shown in
circle markers and its simplified version (algorithm 2) in cross
markers. They have lower performance than Papoutsis' method, but
they increase the range of supported minimum rates. To quantify
these observations, we define the following measurements:
\begin{itemize}
\item $A$: The difference in percentage between the minimum rate
supported by the SUS-based heuristic algorithm
\ref{algo:heur-subc-reass} $r_1$, and Papoutsis' method $r_2$,
i.e., $A= 100 (r_1 - r_2)/r_1$. \item $B$: The difference in
percentage between the minimum rate supported by the SUS-based
heuristic algorithm $r_1$ and its simplified version $r_3$ in
percentage. $A$ and $B$ indicate how much the proposed SUS based
algorithm \ref{algo:heur-subc-reass} increases the range of
supported minimum rates. The larger these measurements are, the
better the proposed algorithm \ref{algo:heur-subc-reass}. \item
$E$: The difference in percentage between the upper dual bound
$u_1$ and the sum rate achieved by Papoutsis' method $u_2$, i.e.
$E= 100 (u_1 - u_2)/u_1$. To compute $u_1$ and $u_2$ we average
the sum rates over the minimum rates supported by Papoutsis'
method. This corresponds to rates between $r_0$ up to $r_2$ in
figure \ref{fig:subc-ass-meth-comp}, where $r_0$ is sum of of the
rates at which the $D$ RT users' rate constraints become active.
We average the sum rates over the same rate interval for all
methods. \item $F$: The difference in percentage between the upper
dual bound and the SUS-based heuristic algorithm
\ref{algo:heur-subc-reass}. \item $G$: The difference in
percentage between the upper dual bound and the simplified
SUS-based heuristic algorithm 2. $E,F$ and $G$ indicate how far
the sum rate is from the upper bound for each method.  The smaller
this measurement is, the better the algorithm.
\end{itemize}

\begin{figure}
\centering
\includegraphics[scale=0.6]{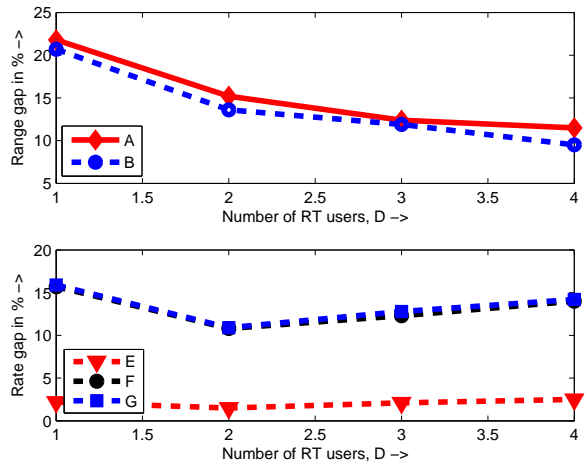}
\caption [Optimal and heuristic methods comparison vs. $D$]
{Optimal and heuristic methods comparison  vs. $D$.}
\label{fig:varL-comp}
\end{figure}

Averaging these measurements over $100$ channel realizations, we
obtain the results shown in figure \ref{fig:varL-comp} for various
number of RT users, $D$. The difference, E, between Papoutsis'
method performance and the upper bound is very small ($<2.5$ \%).
This is because Papoutsis' method minimizes the throughput
reduction by scanning over all possible users swapping. The
proposed heuristic methods have similar performance gaps against
the dual bound ($E,F \approx 13$ \%), which are larger than
Papoutsis' method. However, they achieve this performance with a
much lower computational complexity as we shall see. In addition,
the proposed SUS-based heuristic method with rate-constrained
power allocation, supports up to $20$\% larger minimum rates than
the other two methods (see $A,B$ in figure \ref{fig:varL-comp}).
As the number of RT users $D$ increases, the difference $A$
decreases since it is harder for algorithm
\ref{algo:heur-subc-reass} to find feasible points. Recall that we
force all $D$ user rate constraints to be active.

\begin{figure}
\centering
\includegraphics[scale=0.6]{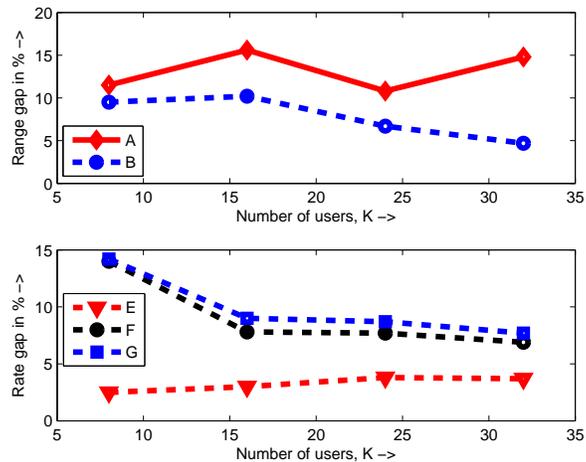}
\caption [Optimal and heuristic methods comparison vs. $K$]
{Optimal and heuristic methods comparison  vs. $K$.}
\label{fig:varK-comp}
\end{figure}

Figure \ref{fig:varK-comp} shows the results when varying the
number of users $K$ from $8$ to $32$. The performance of the
proposed methods improves as the number of users increase, as
indicated by the difference between the upper dual bound and the
sum rate attained (measurements $F$ and $G$ decrease from $14$\%
to $7$\%). This is because in the presence of more users, the SUS
algorithm is more likely to find semiorthogonal channel vectors,
thus increasing the rates and effectively exploiting multiuser
diversity. In contrast, Papoutsis' method slightly deteriorates
when the number of users increase (measurement $E$ increases to
$3.8$ \%).
%

%
\subsection{Increasing the number of RT users} We now study an
interesting scenario where we increase the number of RT users but
keep fixed the minimum rates. This will correspond to the case of
finding the maximum number of supported RT users (e.g. video) in a
cell. We activate the rate constraints by setting the minimum
rates to $10$\% more of the rates achieved by maximum throughput
channel and power allocation.

\begin{figure}
\centering
\includegraphics[scale=0.6]{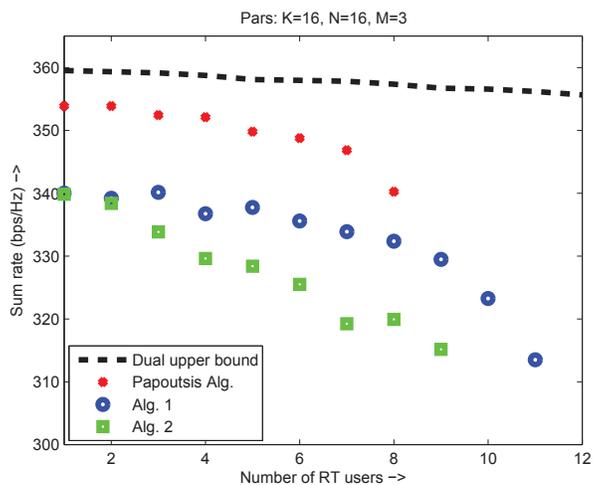}
\caption [Algorithms performance vs. the number of RT users]
{Algorithms performance vs. the number of RT users}
\label{fig:heur-comp-D}
\end{figure}

Figure \ref{fig:heur-comp-D} shows the upper dual bound in dashed
lines, the objective achieved by Papoutsis' method in star markers
and the proposed heuristics in circle and square markers, as a
function of the number of RT users. Papoutsis' method performance
is very close to the upper bound, but it quickly degrades when the
number of RT users increases. It cannot find feasible points with
$9$ RT users onward, while the proposed heuristics yield solutions
for these values within $12$ \% of the upper bound.

\begin{figure}
\centering
\includegraphics[scale=0.6]{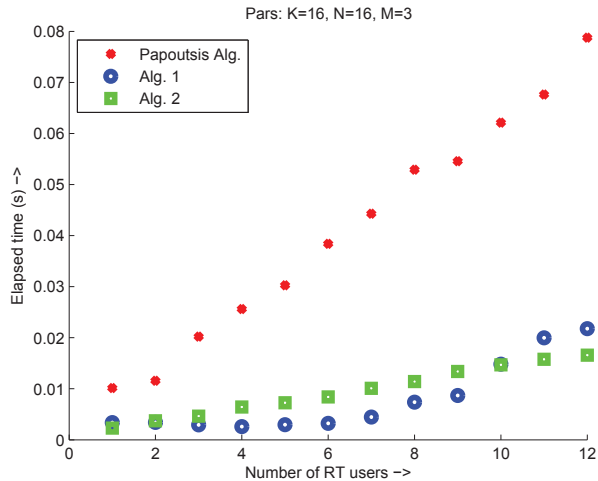}
\caption [Elapsed time vs. the number of RT users] {Elapsed time
vs.  the number of RT users} \label{fig:heur-comp-D-cpu}
\end{figure}

Figure \ref{fig:heur-comp-D-cpu} shows the elapsed time employed
by these heuristics. Papoutsis' method elapsed time grows
approximately linear with the number of RT users and it is larger
than both algorithms \ref{algo:heur-subc-reass} and 2. This is
because when the number of RT users increases,  Papoutsis' method
need to examine more combination of users and invert their
corresponding channel matrices. Algorithms
\ref{algo:heur-subc-reass} and 2 grow much slower. According to
figures \ref{fig:heur-comp-D} and \ref{fig:heur-comp-D-cpu}, not
only algorithms \ref{algo:heur-subc-reass} and 2 support a higher
number of RT users than Papoutsis' method, but they require less
computation time.
%
%
\section{Conclusions} \label{sec:conclusions}
In this paper we designed  algorithms to provide solution points
to the RA problem for ZF beamforming MISO-OFDMA systems supporting
minimum rate requirements.  The solution points given by these
algorithms differ in their distance to the optimal solution and in
the computational complexity to obtain them. We designed two
heuristic methods: algorithms \ref{algo:heur-subc-reass} and 2,
they select an SDMA set for each subchannel and then solve a power
allocation problem.  The difference between algorithms
\ref{algo:heur-subc-reass} and 2 is that the latter one considers
power constraints per subchannel, providing a smaller minimum
rates range but higher computational efficiency.

We showed through numerical evaluations that they have a
performance not far from the optimal solution and that they
increase the range of supported minimum rates when compared with
other approach reported in the literature. This is an important
result because, in a system with RT users, it is more important to
satisfy the rate constraints of the users in need than increasing
the rates of the nRT users. Compared with the method proposed in
\cite{papoutsis11}, our methods do not follow the upper bound as
closely but they increase the range of the minimum rates
supported. This and the fact that we have reduced the
computational complexity, are the main advantages of the proposed
methods. We also showed that rate constrained power allocation
extends the range of the minimum rates supported when applied to
other subchannel assignment methods.

\bibliographystyle{ieeetran}
\bibliography{journal_paper}
%

\end{document}